\documentclass[aps,prb,twocolumn,showpacs,psfig,superscriptaddress,longbibliography]{revtex4-1}
\usepackage{amsfonts}
\usepackage{mathrsfs}
\usepackage{amsmath}
\usepackage{color}
\usepackage{natbib}
\usepackage{textcomp}
\usepackage{graphicx}
\usepackage{bm}
\usepackage{amssymb}

\usepackage{xspace}
\usepackage{epstopdf}
\usepackage{dcolumn}
\usepackage{longtable}
\usepackage{multirow}
\usepackage[colorlinks=true, letterpaper=true, pdfstartview=FitV, linkcolor=blue, citecolor=blue, urlcolor=blue]{hyperref}
\usepackage{float}

\makeatletter

\newcommand{\Rmnum}[1]{\expandafter\@slowromancap\romannumeral #1@}
\makeatother

\begin{document}

\title{Two-dimensional topological superconductivity candidate in van der Waals layered material}

 \author{Jing-Yang You}
 \affiliation{Department of Physics, National University of Singapore, 2 Science Drive 3, Singapore 117551}
 
 \author{Bo Gu}
\affiliation{Kavli Institute for Theoretical Sciences, and CAS Center for Excellence in Topological Quantum Computation, University of Chinese Academy of Sciences, Beijng 100190, China}
 
\author{Gang Su}
\affiliation{Kavli Institute for Theoretical Sciences, and CAS Center for Excellence in Topological Quantum Computation, University of Chinese Academy of Sciences, Beijng 100190, China}

\author{Yuan Ping Feng}
\email{phyfyp@nus.edu.sg}
\affiliation{Department of Physics, National University of Singapore, 2 Science Drive 3, Singapore 117551}
\affiliation{Centre for Advanced 2D Materials, National University of Singapore, 6 Science Drive 2, Singapore 117546}

\begin{abstract}
Two-dimensional (2D) topological superconductors are highly desired because they not only offer opportunities for exploring novel exotic quantum physics, but also possess potential applications in quantum computation. However, there are few reports about 2D superconductors, let alone topological superconductors. Here, we find a 2D monolayer W$_2$N$_3$, which can be exfoliated from its real van der Waals bulk material with much lower exfoliation energy than MoS$_2$, to be a topological metal with exotic topological states at different energy levels. Owing to the Van Hove singularities, the density of states near Fermi level are high, making the monolayer a compensate metal. Moreover, the monolayer W$_2$N$_3$ is unveiled to be a superconductor with the superconducting transition temperature Tc $\sim$ 22 K and a superconducting gap of about 5 meV based on the anisotropic Migdal-Eliashberg formalism, arising from the strong electron-phonon coupling around the $\Gamma$ point, and the 2D superconductor is phonon-mediated and fits the BCS mechanism with Ising-type pairing. Because of the strong electron and lattice coupling, the monolayer displays a non-Fermi liquid behavior in its normal states at temperatures lower than 80 K, where the specific heat exhibit $T^3$ behavior and the Wiedemann-Franz law dramatically violates. Our findings not only provide the platform to study the emergent phenomena in 2D topological superconductors, but also open a door to discover more 2D high-temperature topological superconductors in van der Waals materials.
\end{abstract}
\maketitle

${\color{blue}{Introduction}}$---Two-­dimensional (2D) materials have been extensively studied due to their intriguing properties, such as magnetism~\cite{Huang2017,Gong2017,Deng2018,OHara2018}, topological states~\cite{Bernevig2006,Konig2007,Chang2013}, superconductivity~\cite{Saito2015,Lu2015,Xi2015} and so on. Since the successful exfoliation of graphene in 2004~\cite{Novoselov2004}, several 2D materials including MoS$_2$~\cite{Splendiani2010,Li2011}, CrI$_3$~\cite{Huang2017}, CrGeTe$_3$~\cite{Gong2017}, etc., have been synthesized in experiments. Among these 2D materials, single atomic layer materials are particularly fascinating because they can be easily exfoliated from their van der Waals bulk materials and can be constructed into multifarious heterostructures to realize composite and extraordinary physical phenomena~\cite{Geim2013,Novoselov2016,Baroni2001,Li2016,Zhang2019,Gong2019}.

2D superconductors exfoliated from van der Waals bulk materials represent a unique class of 2D superconductivity because of the easy fabrication and the absence of the substrate~\cite{Saito2016,Guillamon2008,Ugeda2015,Cao2018,Jiang2014,Yu2019,Xi2015a,NavarroMoratalla2016}. While the exotic properties related to them, the discovery of these exfoliated 2D superconductors has been rarely reported. The monolayer transition metal dichalcogenides, NbSe$_2$, TaS$_2$ and TiSe$_2$ display coexisting superconductivity and charge density wave driven by electron-phonon coupling
\cite{Valla2004,Xi2015,Zheng2018,Zheng2019,Ugeda2015,Sipos2008,NavarroMoratalla2016,Klemm2015,Calandra2011}. Recently, the monolayer NiTe$_2$ was reported to be a two-gap superconductivity with a T$_c$ $\sim$ 5.7 K, and the Tc can be enhanced to 11.3 K by inserting one lithium atomic layer into bilayer NiTe$_2$~\cite{Zheng2020}. These 2D superconductors provide great opportunities for exploring fascinating quantum physics. The discovery of 2D topological materials, such as quantum spin Hall~\cite{Bernevig2006,Konig2007}, quantum anomalous Hall insulators~\cite{Chang2013} and topological (semi-) metals~\cite{Burkov2016,Chiu2016,Bansil2016}, sheds insightful light on the study of emerging physical phenomena because of the interplay of superconductivity and nontrivial topology. The emergent topological superconductors with Majorana fermions are thought to be useful for fault-tolerant quantum computation~\cite{Fu2008,Hasan2010,Qi2011}. Topological superconductivity was proposed to be induced in topological boundary states of heterostructures composed of topological materials and superconductors by proximity effects~\cite{Fu2008,Hasan2010,Qi2011}. However, the interface conditions dramatically influence the observation of topological superconductivity. Therefore, discovering 2D materials with simultaneously superconductivity and nontrivial topology will be of great value to achieve topological superconductivity. Although several 2D superconductors have been obtained, very few of them can exhibit topological states~\cite{Lv2017,Liao2018,Menard2017,Fei2017,Leng2017,Liu2018}. Thus, a crucial issue is to fabricate novel 2D materials with simultaneously superconductivity and nontrivial topology, where van der Waals materials may play an important role.

In this work, we report a detail investigation of the superconductivity and nontrivial electronic topology in 2D monolayer W$_2$N$_3$~\cite{Yu2018}. The monolayer W$_2$N$_3$ exhibits a topological metal with three nodal lines traversing the whole Brillouin zone (BZ), which are protected by the symmetries. Topological surface states with respect to these nontrivial band topology are also investigated. Moreover, the monolayer W$_2$N$_3$ is found to be a superconductor with the superconducting transition temperature of about 22 K and the superconducting gap up to 5 meV based on the anisotropic Midgal-Eliashberg formalism. This 2D superconductor is phonon-mediated and fits the BCS mechanism with Ising-type pairing. The high Tc and large superconducting gap are revealed to result from the enhanced electron-phonon coupling (EPC) near Fermi level at $\Gamma$ point. The strong electron-phonon coupling come from the high density of states near Fermi level brought about by the Van Hove singularities near Fermi level, leading to more electrons susceptible to pairing mediated by phonons. Besides, owing to the strong electron-phonon coupling, the normal states of monolayer W$_2$N$_3$ perform as a non-Fermi liquid at temperatures lower than 80 K, where the electrical specific heat shows a cubic dependence of temperature and the Wiedemann-Franz law is obviously violated. The coexistence of superconductivity and nontrivial band topology in this layered material makes it an inevitable platform to study topological superconductivity. 

${\color{blue}{Calculation \ Method}}$---Our first-principles calculations were based on the density functional theory (DFT) as implemented in the QUANTUM ESPRESSO package~\cite{Giannozzi2009}, using the full relativistic pseudopotential. The vacuum layer was set to 15 \AA. To warrant an energy convergence of less than 1 meV per atom, the plane waves kinetic-energy cutoff was set as 100 Ry and the energy cutoff for charge density was set as 1250 Ry. The structural optimization was performed until the forces on atoms were less than 1 meV/Å. An unshifted BZ $\mathbf{k}$ point mesh of 20$\times$20$\times$1 was utilized for electronic charge density calculations. The phonon modes are computed within density-functional perturbation theory~\cite{Baroni2001} on a 10$\times$10$\times$1 $\mathbf{q}$ mesh. The EPW code~\cite{Giustino2007,Margine2013,Ponce2016} was employed for the calculation of superconducting gap with both fine $\mathbf{k}$ and $\mathbf{q}$ meshes of 100$\times$100$\times$1 in the BZ. The surface spectrum was calculated by using the Wannier functions and the iterative Green's function method~\cite{Marzari1997, Souza2001, Wu2018, Sancho1985}. The electronic and phonon transport properties were calculated with the packages BoltzTrap~\cite{Madsen2006} and ShengBTE~\cite{Li2014}, respectively.

\begin{figure}[!htbp]
  \centering
  \includegraphics[scale=0.28,angle=0]{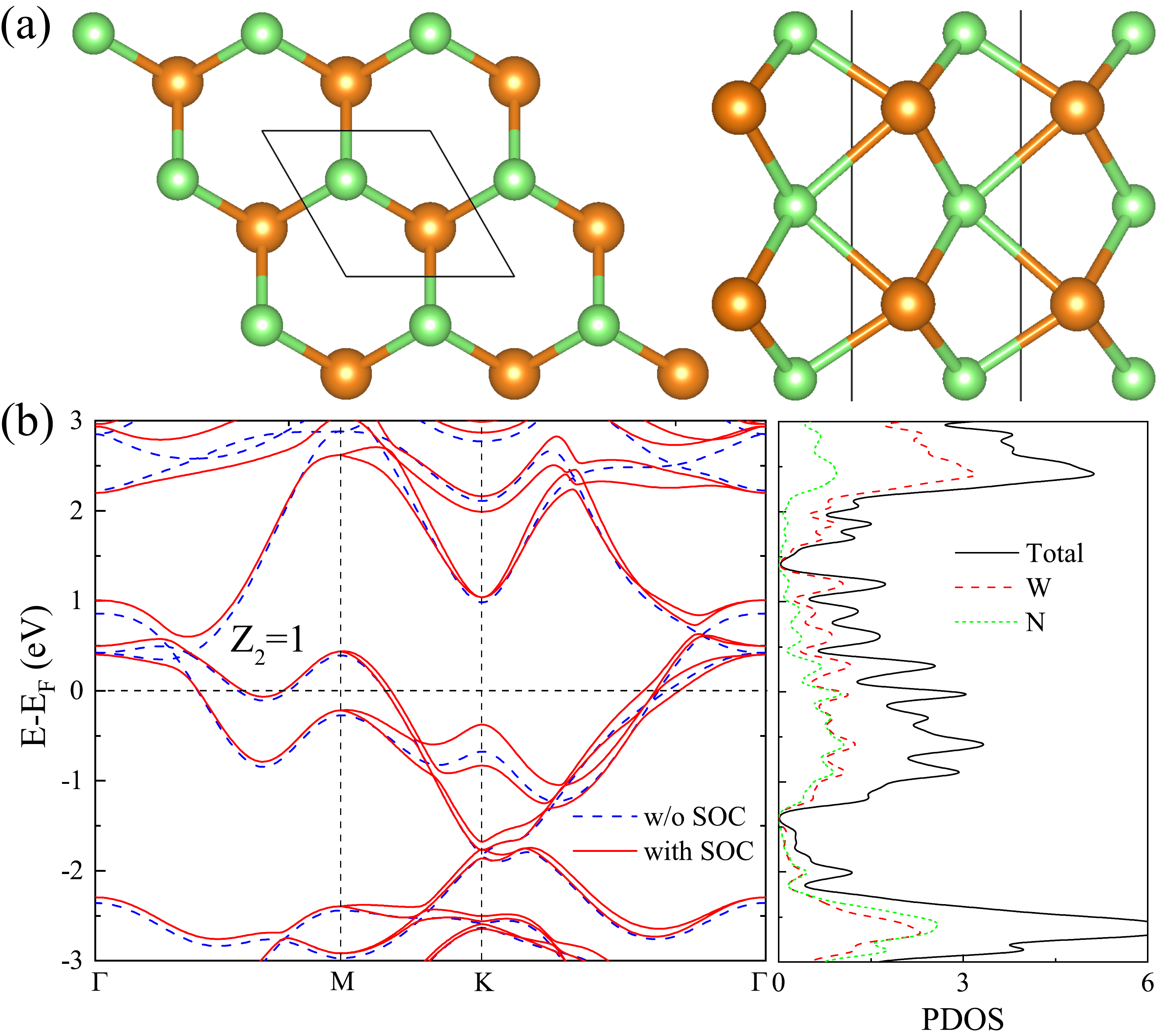}\\
  \caption{(a) Top and side views of the crystalline structure of monolayer W$_2$N$_3$. (b) Electronic band structures of monolayer W$_2$N$_3$ without (dash blue lines) and with (solid red lines) SOC, and the corresponding partial density of states (DOS) with SOC. The Brillouin zone (BZ) with high symmetry paths is inserted.}\label{fig1}
\end{figure}

${\color{blue}{Crystal \ structure}}$---The bulk W$_2$N$_3$ has a van der Waals layered crystal structure with the space group of P6$_3$/mmc (No.194)~\cite{Wang2012}. It is composed of inversion symmetric N-W-N-W-N layers with two 1T N-W-N layers sharing one layer of N atoms as shown in Fig.~\ref{fig1}(a), and shows AB stacking along the (0001) direction. The monolayer W$_2$N$_3$ with the space group of P$\bar{6}$m2 (No.187) can be easily exfoliated from its bulk material due to the low exfoliation energy of 46.6 meV/atom~\cite{Yu2018,Mounet2018}, which is much lower than that for monolayer MoS$_2$ with 76.3 meV/atom~\cite{Zhou2019}. The in-plane lattice constant of monolayer W$_2$N$_3$ is optimized to be $a = 2.864$ \AA, which is reasonable compared with the in-plane lattice constant of its bulk~\cite{Wang2012}. 

\begin{figure}[!htbp]
  \centering
  \includegraphics[scale=0.3,angle=0]{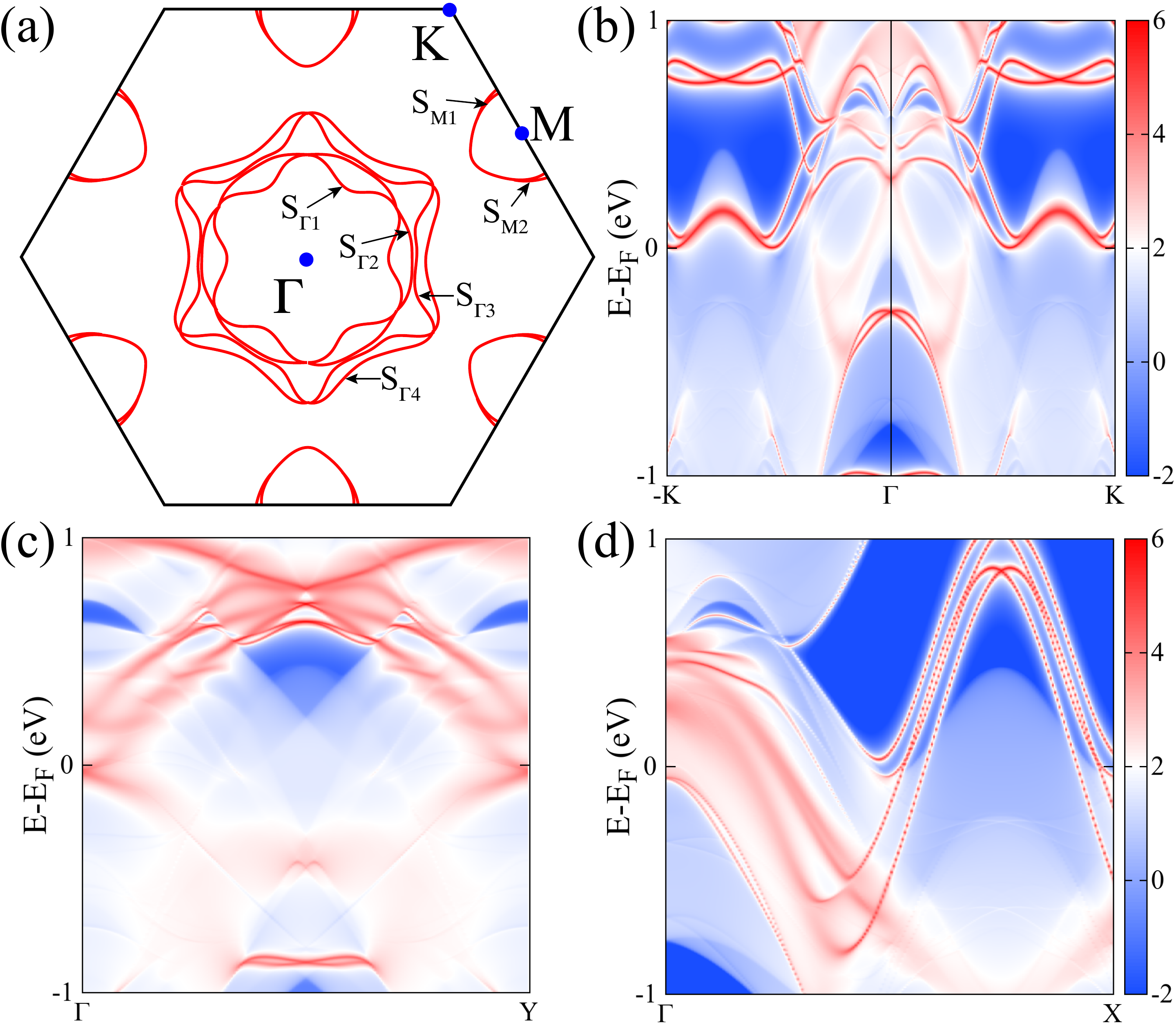}\\
  \caption{(a) Fermi surfaces (red lines) for monolayer W$_2$N$_3$.  The surface states of monolayer W$_2$N$_3$ projected on (b) (111), (c) (100) and (d) (010) planes, respectively. Warmer colors represent higher local density of states, and blue regions indicate the bulk band gap.}\label{fig2}
\end{figure}

${\color{blue}{Topological \ states}}$---The electronic structure of monolayer W$_2$N$_3$ without and with spin-orbital coupling (SOC) as well as the partial DOS with SOC are plotted in Fig.~\ref{fig1}(b). By comparing the electronic band structures of monolayer W$_2$N$_3$ without and with SOC, one may observe that the SOC removes some degenerate points and significantly changes the band structure because of the heavy W atoms possessing relative large SOC. Thus, all results discussed in the following were calculated with SOC included. For monolayer W$_2$N$_3$,  there are four bands crossing the Fermi level, making the monolayer a compensated metal as shown in Fig.~\ref{fig1}(b). There are three flower-shaped hole pockets (labeled as $S_{\Gamma_{1,2,3}}$, from the inside out) and one electron pocket ($S_{\Gamma_{4}}$) at $\Gamma$ point. At the centers of the boundary of BZ, M, are elliptical hole ($S_{{\rm M}_{1}}$) and electron ($S_{{\rm M}_{2}}$) pockets [Fig.~\ref{fig2}(a)]. The W and N atoms contribute almost equally to the low energy band structure and DOS near Fermi level because of the large $p$-$d$ hybridization. The band structures present many interesting features: the SOC lifts the degenerate points at 0.5 eV above Fermi level, leading to the nontrivial band topology characterized by the topological invariant Z$_2$ = 1; both without and with SOC the Weyl points at about 1.5 eV below Fermi level at K (K$^\prime$) points are stable, which are protected by the symmetries at K (K$^\prime$) points; it is interesting to note that the two-fold band degeneracy along the high symmetry lines $\Gamma$-M exists both without and with SOC; there are several Van Hove singularities in the band structure leading to sharp peaks of DOS, in particularly the Van Hove singularities at about 0.1 eV below Fermi level bringing about high DOS near Fermi level. The two-fold degeneracy along $\Gamma$-M high symmetry lines is protected by the symmetries, because when SOC is included any point on $\Gamma$-M lines belongs to the group $G^5_8$ with the reality of $a$, which only possesses two-dimensional irreducible representation~\cite{Wondratschek1973}, constraining the two-fold degeneracy along $\Gamma$-M at all energy scale. These nontrivial topological states are fully reflected in their surface states as shown in Figs.~\ref{fig2}(b), (c) and (d), where abundant topological surface states can be seen.

\begin{figure}[!htbp]
  \centering
  \includegraphics[scale=0.25,angle=0]{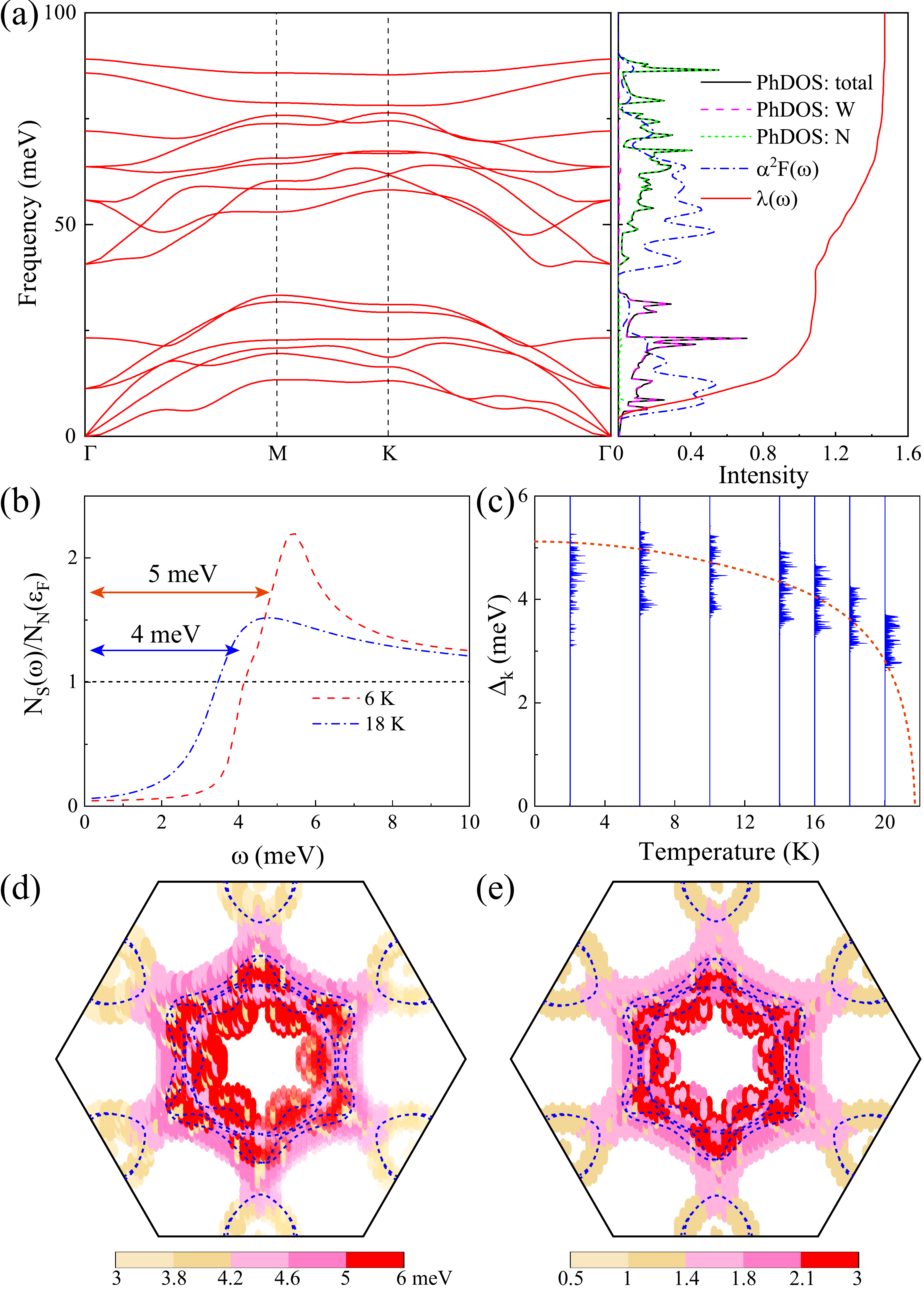}\\
  \caption{ (a) The phonon spectrum and phonon density of states (PhDOS) (multiply 5), Eliashberg spectral function
$\alpha^2F(\omega)$, and the cumulative frequency-dependent of EPC $\lambda(\omega)$ of monolayer W$_2$N$_3$ at ambient pressure. (b) Quasiparticle DOS in the superconducting state for two representative temperatures of 6 K (red dashed line) and 18 K (blue dash-dotted line). The black dashed line indicates the DOS of normal state, which is normalized to 1 at the Fermi level. The superconducting DOS $N_S(\omega)$ becomes 1 beyond the highest phonon energy. (c) Anisotropic superconducting gap of monolayer W$_2$N$_3$ on the Fermi surface as a function of temperature. Red dashed curve is BCS fit of the superconducting gap. (d) Momentum-resolved superconducting gap $\Delta_k$ on the Fermi surface (blue dashed lines) at 10 K, where warmer color (red) indicates bigger $\Delta_k$. (e) Momentum-resolved electron-coupling strength $\lambda_k$ within 0.2 eV on the Fermi surface (blue dashed curves). Warmer colors (dark red) indicates larger $\lambda_k$.}\label{fig3}
\end{figure}

${\color{blue}{Superconductivity}}$---The metallic property with high DOS near Fermi level motivated us to investigate the possible superconductivity. The phonon spectra, the phonon density of states (PhDOS), the Eliashberg electron–phonon spectral function $\alpha^2F(\omega)$ and the cumulative frequency dependent EPC $\lambda(\omega)$ are shown in Fig.~\ref{fig3}(a). The absence of imaginary frequency modes indicates the dynamical stability of monolayer W$_2$N$_3$ as shown in Fig.~\ref{fig3}(a). From the projected PhDOS, we find that W atoms vibrate in the low frequency region, while N atoms vibrate in the relative higher frequency region because their distinct atomic masses, and the phonon vibrational modes of W and N atoms are separated by a phonon spectral energy gap of about 6 meV. $\alpha^2F(\omega)$ displays a dominant peak centered around 10 meV at the low frequency region, while $\alpha^2F(\omega)$ is more spread at the high frequency region. The cumulative frequency-dependent EPC $\lambda(\omega)$ can be calculated by integrating $\alpha^2F(\omega)$. From the calculated $\lambda(\omega)$, we find that the low-frequency phonons (below the phonon spectral gap (35 meV)) account for 74$\%$ of the total EPC [$\lambda$ = $\lambda(\infty)$ = 1.47]. Thus, W atoms make the main contribution to the EPC. Utilizing our calculated $\alpha^2F(\omega)$ and $\lambda(\omega)$ together with a typical value of the effective screened Coulomb repulsion constant $\mu^\ast$ = 0.1, the superconducting transition temperature T$_c$ is calculated to be 21 K with the McMillan-Allen-Dynes (MAD) approach~\cite{McMillan1968,Allen1975,Giustino2017}, which is highest reported for 2D superconductors exfoliated from van der Waals bulk materials.

Figure~\ref{fig3}(b) shows the quasiparticle DOS in the superconducting state at two representative temperatures of 6 and 18 K, relative to the DOS in the normal state as a function of frequency $\omega$ (meV) based on the anisotropic Migdal-Eliashberg theory~\cite{Giustino2007,Margine2013,Ponce2016}. We can observe that the superconducting gaps at 6 and 18 K are about 5 and 4 meV, respectively. The $k$-resolved superconducting gaps on the Fermi surface, $\Delta_k(T)$, for several temperatures below 20 K with $\omega$=0, as well as the BCS fit for the superconducting gap are display in Fig.~\ref{fig3}(c). It is seen that the superconducting gap vanishes at around 21.7 K, which is a little higher than the T$_c$ (21 K) obtained with McMillan-Allen-Dynes (MAD) approach, and the zero-temperature superconducting gap is about 5 meV.

Figure~\ref{fig3}(d) shows the momentum-resolved superconducting gap on the Fermi surface. On the whole Fermi surface, there are finite superconducting gaps, indicating the Fermi surface is fully gapped below T$_c$. The superconducting gap on the $S_{\Gamma_1}$ line exhibits larger gap distribution around the lowland on $\Gamma$-K paths, while the largest superconducting gaps on the $S_{\Gamma_3}$ and $S_{\Gamma_4}$ lines distribute on $\Gamma$-M paths, and the superconducting gap on the $S_{\Gamma_2}$ line is smaller than that on the above three Fermi surface lines. The superconducting gaps on the $S_{{\rm M}_1}$ and $S_{{\rm M}_2}$ lines is about 2 meV smaller than that on the $S_{\Gamma}$ lines. It is noted that the superconducting gaps both on the $S_{\Gamma}$ and $S_M$ lines are highly anisotropic. The regions of the Fermi surface with the largest superconducting gap coincide with the electron-phonon coupling strength $\lambda_k$ as shown in Fig.~\ref{fig3}(e). Thus, the high Tc and large $\Delta_k$ in monolayer W$_2$N$_3$ are mainly from the strong electron-phonon coupling of Fermi surface at $\Gamma$ point. 

\begin{figure}[!!!!htbp]
  \centering
  \includegraphics[scale=0.3,angle=0]{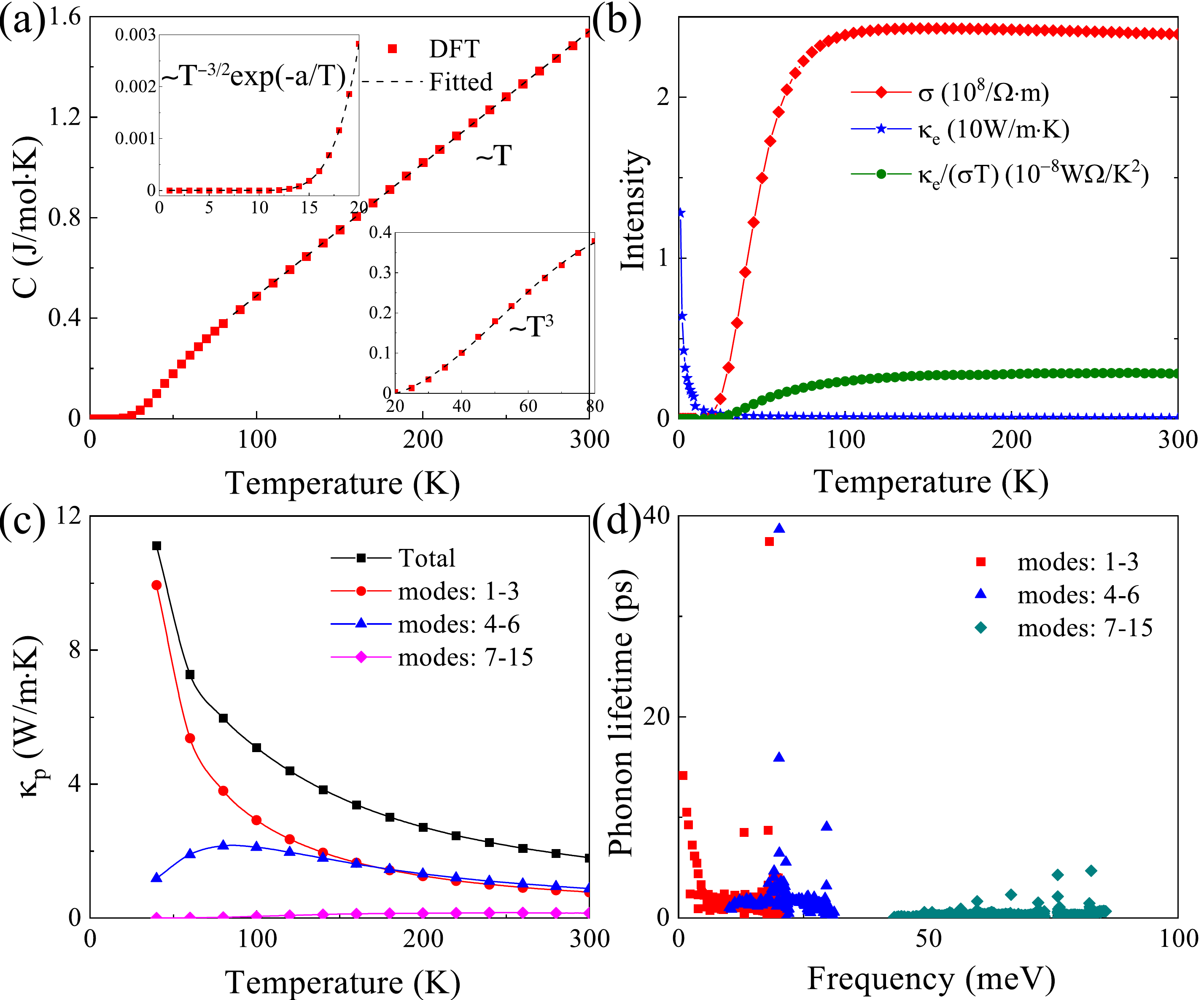}\\
  \caption{(a) Temperature dependence of electronic specific heat C of monolayer W$_2$N$_3$. The upper and lower insets are the enlarged parts of the specific heat at temperatures below T$_c$ (T $<$ 20 K), and above T$_c$ (20 $<$ T $<$ 80 K), respectively. The black dashed lines are fitting curves at different temperature regions. (b) Temperature dependent electrical ($\sigma$) and thermal ($\kappa_e$) conductivities, and Lorenz number $L$ [$L=\kappa_e/(\sigma T)$]. (c) The modes-resolved lattice thermal conductivity $\kappa_p$ as a function of temperature. (d) The modes-resolved phonon lifetime as a function of frequency at 300 K.}\label{fig4}
\end{figure}

${\color{blue}{Transport \ properties}}$---It is worth mentioning that due to the exceedingly strong coupling between the electrons and phonons, the normal state of monolayer W$_2$N$_3$ may be a non-Fermi liquid. The temperature dependent electronic specific heat of monolayer W$_2$N$_3$ is calculated as shown in Fig.~\ref{fig4}(a). It is noted that the specific heat C(T) displays distinct features: at low temperature (below Tc), $C(T) \sim T^{-3/2}\cdot exp(-a/T)$, i.e. the specific heat accords with the low-temperature specific heat of BCS superconductors (upper inset);  at 20 $<$ $T$ $<$ 80 K, $C(T) \sim T^3$ (lower inset); and at high temperature region ($T$ $>$ 80 K), $C(T) \sim T$. It is clear that below 80 K, the normal stat shows a non-Fermi liquid behavior~\cite{Schofield1999}.

To further verify this observation, the temperature dependent Lorenz number $L$ [=$\kappa_e/(\sigma T)$] is also studied, as  presented in Fig.~\ref{fig4}(b). We can observe that the Lorenz number $L$ is not a constant at temperature below 80 K, which dramatically violates the Wiedemann-Franz law~\cite{Franz1853}, while at high temperature it shows almost a constant, indicating a Fermi liquid behavior. 

The lattice thermal conductivity is given in Fig.~\ref{fig4}(c). It can be observed at temperatures lower than 200 K, the lattice thermal conductivity dominates, while at temperatures above 200 K, the electrical thermal conductivity dominates. It is noted that at low temperatures, the three acoustic modes (modes: 1-3) make the main contribution to lattice thermal conductivity, while with the increase of temperature the contribution of three optical branches from the vibration of N atoms (modes: 4-6) begins to match or even surpass that of the acoustic branches. The optical branches from vibration of N atoms (modes: 7-15) make little contribution to the thermal conductivity at the whole temperature range. 

To gain more insight into phonon transport of monolayer W$_2$N$_3$, the modes-resolved phonon lifetime is plotted in Fig.~\ref{fig4}(d). It is found that most of phonon lifetimes of three acoustic branches are comparable with three optical branches from W atoms vibration, showing a comparable contribution to the lattice thermal conductivity, while the phonon lifetime of the optical branches above 40 meV is shorter because of the larger scatting rate. Thus, the N atoms contribute little to the thermal conductivity, leading to the low thermal conductivity of monolayer W$_2$N$_3$.

\begin{figure}[!!!!htbp]
  \centering
  \includegraphics[scale=0.5,angle=0]{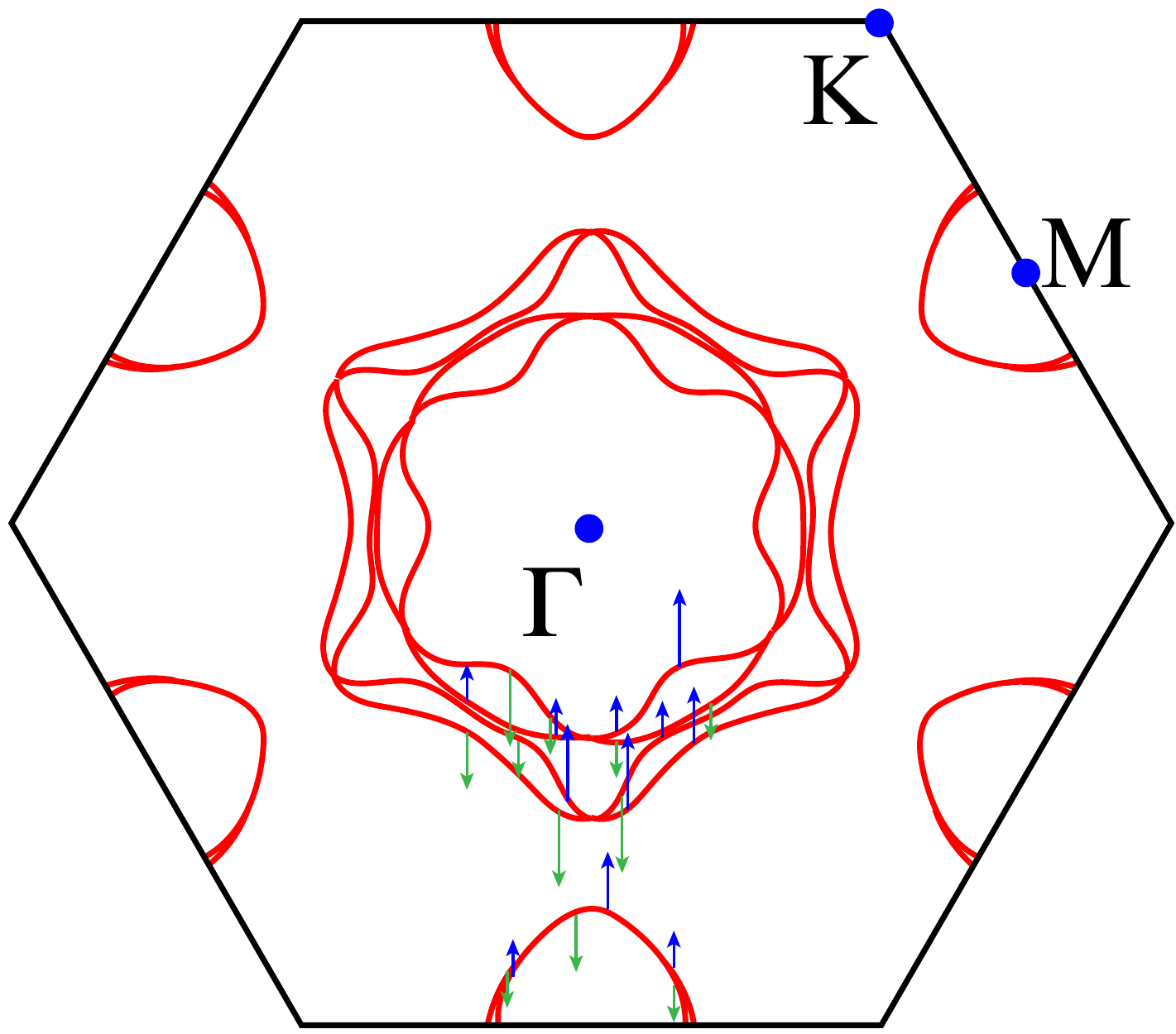}\\
  \caption{Spin texture of electronic bands that cross the Fermi level, where the up (blue) and down (green) arrows represent the opposite out-of-plane spin polarization directions. In this figure we only plot the spin texture of partial Fermi surface, and the spin texture in other parts of the Fermi surface can be obtained by symmetries.}\label{fig5}
\end{figure}

{\color{blue}\em Discussion}---To unveil the pairing mechanism and possible relation between the nontrivial band topology and superconductivity, the spin texture of Fermi surface is calculated as plotted in Fig.~\ref{fig5}. From Fig.~\ref{fig5}(a), it is noted that each Fermi line is fully out-of-plane spin polarized, indicating an Ising-type pairing in monolayer W$_2$N$_3$. This result is similar to the superconductivity in ion-gated MoS$_2$~\cite{Saito2015}. However, the magnitude of spin moments is different at different $\mathbf{k}$ points on Fermi surface for monolayer W$_2$N$_3$. The spin polarization direction between the outer and inner Fermi surface is inverted at the degenerate points on $\Gamma-M$ paths, at which the spin up and spin down are cancelled. Comparing the Figs.~\ref{fig3}(d) and \ref{fig5}, it is interesting to note that the strength of electron-phonon coupling is related to the magnitude of spin moment, that is to say, where the magnitude of spin moment is large, the electron-phonon coupling is also enhanced. Since the superconductivity is related to the degeneracy point, which determines the topological metal state of monolayer W$_2$N$_3$, there may be some correlation between the topology and superconductivity in our system, which needs further deep-going study. 

${\color{blue}{Summary}}$---In this work, we report that a 2D monolayer W$_2$N$_3$ hosts simultaneously topological states and superconductivity. The monolayer is found to exhibit different topological states at different energy level including topological Z$_2$ insulator, Weyl semimetal and topological nodal lines metal. The exotic topological surface states are also investigated. Furthermore, based on anisotropic Migdal-Eliashberg theory, the monolayer W$_2$N$_3$ is predicted to be a phonon-mediated BCS superconductor with the superconducting transition temperature Tc $\sim$ 22 K and the superconducting gap $\Delta_k \sim 5$ meV. The high Tc and large $\Delta_k$ are unveiled to be from the enhanced electron and lattice coupling near Fermi level at $\Gamma$ point. Due to the large electron-phonon coupling, the normal state of monolayer W$_2$N$_3$ performs as a non-Fermi liquid at temperatures below 80 K, where the electrical specific heat displays a T$^3$ behavior and the Wiedemann-Franz law is dramatically violated. Our results will spur the observation of 2D high-temperature topological superconductivity.

\section*{Acknowledgement}
This research/project is supported by the Ministry of Education, Singapore, under its MOE AcRF Tier 3 Award MOE2018-T3-1-002. B.G. and G.S. are supported in part by the National Key R\&D Program of China (Grant No. 2018YFA0305800), 
the Strategic Priority Research Program of the Chinese Academy of Sciences (Grant No. XDB28000000), and NSFC (Grant No.11834014).


%

\end{document}